\documentclass[reprint,amsmath,amssymb,aps,prc,superscriptaddress]{revtex4-2}
\usepackage{graphicx}
\usepackage{epstopdf}
\usepackage{dcolumn}
\usepackage{bm}
\usepackage[justification=Justified,font=smaller]{caption}
\usepackage{booktabs}
\usepackage{hyperref}
\usepackage{color}
\hypersetup{hypertex=true,
			colorlinks=true,
			linkcolor=blue,
			anchorcolor=blue,
			citecolor=blue,
			}
\usepackage[labelformat=simple]{subcaption}

\begin{document}


\title{Effects of phase transition in hybrid stars from quark-meson coupling hadronic matter to deconfined quark matter}


\author{Yida Yang}
\affiliation{School of Physics and Electronic Science, East China Normal University, Shanghai 200241, China\\}
\author{Chen Wu}
\affiliation{Shanghai Institute of Applied Physics, Chinese Academy of Sciences, Shanghai 201800, China\\}
\author{Ji-Feng Yang}\thanks{Corresponding author}
\affiliation{School of Physics and Electronic Science, East China Normal University, Shanghai 200241, China\\}

\date{\today}

\begin{abstract}
 Different types of phase transition from hadron to quark at high density near zero temperature may occur in the inner core of hybrid stars. We investigate the impacts of phase transition types and quark models on properties of hybrid star and quark cores. The quark-meson coupling (QMC) model is used to describe hadronic matter, and the MIT bag model as well as the Nambu-Jona-Lasinio (NJL) model are employed to describe quark matter for comparison. From the mass-radius curves obtained by using equations of state (EOS), we find that EOSs of hadron matter have a decisive influence on the maximum mass of a hybrid star in the first-order phase transition case, while quark matter EOSs have more impacts on results in the crossover transition case. It is also found in the present work that the thermodynamic correction arising from an interpolation scheme considerably stiffens the EOSs. Therefore the crossover type phase transition generally leads to hybrid stars with higher masses. In particular, by using the QMC model and \text{the} NJL model to construct crossover EOSs with thermodynamic correction, we discover that the maximum masses of hybrid stars can meet the recent observational constraint on mass from PSR J0952-0607, i.e., $2.35\pm0.17M_\odot$.
\end{abstract}

\maketitle

\section{\label{sec1}INTRODUCTION}
Neutron stars are one of the end points in the evolution of massive stars and are composed of dense matter, in which the density in the inner core could reach several times the nuclear saturation density ($\rho_{0}\approx0.15~\mathrm{fm}^{-3}$) \cite{ozelMassesRadiiEquation2016,maQuarkDeconfinementNeutron2007,pandaHybridStarsQuarkmeson2004}. The study of neutron stars is important to our understanding of dense matter properties. Although relativistic heavy-ion collisions \cite{gyulassyNewFormsQCD2005} have advanced our understanding of hot dense QCD matter in the last few decades, neutron stars are the only known natural laboratories containing cold and dense QCD matter \cite{baymHadronsQuarksNeutron2018,masudaHadronQuarkCrossover2013,liHadronQuarkCrossover2022,maKaonMesonCondensate2022,yinSlowlyRotatingNeutron2010}. 

It has long been thought that QCD matter undergoes a phase transition from hadronic matter to deconfined quark matter at high densities and/or high temperatures due to the asymptotically freedom of the strong interactions \cite{xiaSystematicStudyQuarkhadron2020,karimiHybridStarsFramework2023,sorensenPhaseTransitionsCritical2021,fukushimaPhaseDiagramDense2011,vanheugtenFermiliquidTheoryImbalanced2012,PhysRevLett.30.1343}, and that such a phase transition is very likely to take place in the high-density regions of neutron star \cite{collinsSuperdenseMatterNeutrons1975,somasundaramInvestigatingSignaturesPhase2023,sedrakianHeavyBaryonsCompact2023}, especially since some model-independent calculations in recent years have suggested that the presence of quark cores inside massive neutron stars should be considered the standard scenario \cite{annalaEvidenceQuarkmatterCores2020}. Such neutron stars with both hadronic matter and deconfined quark matter are called hybrid stars, and the research of hybrid stars can help us further understand the phase transition properties of dense QCD matter in the low-temperature region.

In order to describe hadron-quark phase transitions near zero temperature, one need to obtain equations of state (EOS). Ideally, it would be preferable to calculate the EOS directly from QCD and extract information about the phase transitions. Unfortunately, there is no reliable first principle QCD calculation that can produce an equation of state applicable in high density regions  \cite{whittenburyHybridStarsUsing2016,liStructuresStrangeQuark2019,klahnModernCompactStar2007,wuEffectsQuarkmatterSymmetry2018,chang-qunPhaseTransitionsNeutron2008}. Thus one is forced to employ phenomenological models to characterize hadronic matter and quark matter separately and connect their EOSs via proper means.

In this work, we use the quark-meson coupling (QMC) model \cite{GUICHON1988235,rikovskastoneColdUniformMatter2007,whittenburyQuarkmesonCouplingModel2014,guichonQuarkMesonCouplingQMC2018} to describe the hadronic phase, and the quark phase will be described by the MIT bag model \cite{chodosNewExtendedModel1974} and the NJL model \cite{nambuDynamicalModelElementary1961,hatsudaQCDPhenomenologyBased1994,buballaNJLmodelAnalysisDense2005} that are 'almost complementary' \cite{buballaNJLmodelAnalysisDense2005}. We hope the EOSs thus obtained to be more reliable in the phase transition region \cite{whittenburyHybridStarsUsing2016}. For the connection between the hadronic and quark phases, we will follow the main treatments in literature \cite{baymHadronsQuarksNeutron2018,fukushimaPhaseDiagramDense2011,drischlerChiralEffectiveField2021,ranea-sandovalEffectsHadronQuarkPhase2019}: One is the first-order phase transition described by Maxwell \cite{karimiHybridStarsFramework2023,oertelEquationsStateSupernovae2017,lenziHybridStarsReactive2023,maieronHybridStarsColor2004,alaverdyanQuarkMatterNJL2022,contreraQuarknuclearHybridEquation2022} or Gibbs \cite{pandaHybridStarsQuarkmeson2004,karimiHybridStarsFramework2023,masudaHyperonPuzzleHadronquark2016,xiaSystematicStudyQuarkhadron2020,wuEffectsQuarkmatterSymmetry2018,juHadronquarkMixedPhase2021,yangInfluenceHadronicEquation2008,chang-qunPhaseTransitionsNeutron2008,liuPropertiesQuarkmatterCores2023,schertlerQuarkPhasesNeutron2000} constructions, another is the crossover \cite{masudaHadronQuarkCrossover2013,whittenburyHybridStarsUsing2016,masudaHyperonPuzzleHadronquark2016,liHadronQuarkCrossover2022,masudaHADRONQUARKCROSSOVERMASSIVE2013,liPropertiesHybridStars2022} type. Our aim is to study the effects of different quark models and phase transitions on the maximum masses of hybrid stars. 
Recent and earlier observations of massive neutron stars, PSR J0952-0607 (2.35 ± 0.17 times the solar mass $M_{\odot}$) \cite{romaniPSRJ095206072022} and PSR J0348+0432 (2.01 ± 0.04$M_{\odot}$) \cite{antoniadisMassivePulsarCompact2013}, pose challenges to the traditional understanding of the EOSs of super-dense matter\cite{maKaonmesonCondensationResonance2023,juHadronquarkMixedPhase2021}. We wish to see whether our calculation could yield masses compatible with these data. In addition, we will also study the properties of quark cores in hybrid stars.

This paper is organized as follows. In Sec.~\ref{sec2}, the QMC model with hyperons is presented. In Sec.~\ref{sec3}, we briefly introduce the MIT Bag model and \textbf{the} SU(3) NJL model. Combining the above models we obtain the hybrid equation of state by using the Gibbs criterion and interpolation scheme in Sec.~\ref{sec4}. Our calculations of the hybrid star properties are given and analyzed in Sec.~\ref{sec5}. Included in Sec.~\ref{sec6} are our conclusions and summaries.
\section{\label{sec2}HADRONIC MATTER DESCRIPTION WITHIN THE QMC MODEL}
In the QMC model, a baryon immersed in the nuclear medium is treated as a static MIT bag containing quarks. The interactions between baryons are realized by exchanging scalar ($\sigma$) and vector ($\omega,\rho$) mesons which are regarded as classic fields in the mean-field approximation and couple directly to the quarks in the bag. The equation of motion of quark fields inside the bag reads: 
\begin{eqnarray}
	\left[i\gamma^{\mu}\partial_{\mu}-\big(m_{q}-g_{\sigma}^{q}\sigma\big)-\gamma^{0}\big(g_{\omega}^{q}\omega+g_{\rho}^{q}\frac{\tau_{3}}{2}\rho\big)\right]\psi_{q}=0,
\end{eqnarray}
where $q=u,d,s$, with the current quark mass $m_{q}$ and the quark-meson coupling constants $g_{\sigma}^{q},g_{\omega}^{q},g_{\rho}^{q}$. The normalized ground state for a quark in the bag is given by
\begin{eqnarray}
	\psi_q(\mathbf{r},t)=\mathcal{N}_qe^{-i\epsilon_qt/R_{B}}\bigg(\begin{matrix}j_0(x_qr/R_{B})\\i\beta_q\sigma\cdot\hat{\mathbf{r}}j_1(x_qr/R_{B})\end{matrix}\bigg)\frac{\chi_q}{\sqrt{4\pi}},
\end{eqnarray}
where
\begin{eqnarray}
	&&
	\epsilon_{q}=\Omega_{q}+R_{B}\big(g_{\omega}^{q}\omega+g_{\rho}^{q}\frac{\tau_{3}}{2}\rho\big),\\
	&&
	\beta_q=\sqrt{\frac{\Omega_q-R_{B}m_q^*}{\Omega_q+R_{B}m_q^*}},
\end{eqnarray}
with $\Omega_{q}=\sqrt{x_{q}^{2}+(R_{B}m_{q}^{*})^{2}}$ and the effective mass of quark $m_{q}^{*}=m_{q}-g_{\sigma}^{q}\sigma$. Here $R_{B}$ and $\chi_{q}$ denote the bag radius and quark spinor respectively. The bag eigenvalue $x_{q}$ is determined by the boundary condition at the bag surface
\begin{eqnarray}
 	j_{0}(x_{q})=\beta_{q}j_{1}(x_{q}).
\end{eqnarray}
The energy of a static bag containing three ground state quarks of different baryons is then given by
\begin{eqnarray}
	E_{B}^{\mathrm{bag}}=\sum_{q}n_{q}\frac{\Omega_{q}}{R_{B}}-\frac{Z_{B}}{R_{B}}+\frac{4}{3}\pi R_{B}^{3}B_{B},
\end{eqnarray}
where the parameter $Z_{B}$ accounts for the zero-point 
motion and center-of-mass corrections and $B_{B}$ is the bag constant. The effective nucleon mass at rest in the model is taken to be $M_{B}^{*}=E_{B}^{\mathrm{bag}}$, and the bag radius $R_{B}$ is determined by minimizing the effective nucleon mass, i.e., the  stability condition
\begin{eqnarray}
	\frac{dM_B^*}{dR_B}=0.
\end{eqnarray}
In order to fix the parameters above, we start from fixing $B_{N}$ and $Z_{N}$ for nucleon, here we set $R_{N}=0.6\mathrm{fm}$, then they are obtained by fitting the nucleon mass $M_{N}=939~\mathrm{MeV}$ and enforcing the stability condition. With the current quark masses $m_{u}=m_{d}=5.5~\mathrm{MeV},~m_{s}=150~\mathrm{MeV}$, we obtained $Z_{N}=4.00506,~B_{N}^{1/4}=210.854~\mathrm{MeV}$. For the remaining baryons, their bag constants are fixed to $B_{B}^{1/4}=210.854~\mathrm{MeV}$, then by performing the similar procedure, $R_{B}$ and $Z_{B}$ of all octet baryons are determined, and the results are displayed in Table~\ref{table1}
\begin{table}[h]
	\captionsetup{singlelinecheck=false}
	\caption{\label{table1}
	Values of $R_{B}$ and $Z_{B}$ for different octet baryons fitting to their physical masses, with the bag constant $B_{B}^{1/4}=210.854\mathrm{~MeV}$, current quark masses $m_{u}=m_{d}=5.5\mathrm{~MeV}\mathrm{~and~}m_{s}=150\mathrm{~MeV}$.}
	\renewcommand{\arraystretch}{1.3}
	\begin{ruledtabular}
		\begin{tabular}{cccc}
			Baryons&
			$M_B$&
			$R_{B}$&
			$Z_{B}$\\
			\hline
			N&			939.0&	0.6&	 4.00506\\
			$\Lambda$&	1115.6&	0.63188& 3.69130\\
			$\Sigma^+$&	1189.3&	0.64577& 3.45691\\
			$\Sigma^0$&	1192.5&	0.64630& 3.44661\\
			$\Sigma^-$&	1197.4&	0.64726& 3.43082\\
			$\Xi^0$&	1314.9&	0.66248& 3.29602\\
			$\Xi^-$&	1321.3&	0.66359& 3.27512\\
		\end{tabular}
	\end{ruledtabular}
\end{table}

The next step is to fit the quark-meson coupling constants $g_{\sigma}^{q},~g_{\omega}^{q}$ and $g_{\rho}^{q}$. Here we take the values of these coupling constants from Ref.~\cite{pandaHybridStarsQuarkmeson2004}:
\begin{eqnarray}
 &&
 g_{\sigma}^{q}=5.957,~g_{\omega N}=8.981,~g_{\rho N}=8.651,\nonumber
 \\
 &&
 g_{\omega}^{q}=\frac{1}{3}g_{\omega N},~g_{\rho}^{q}=g_{\rho N},\nonumber
\end{eqnarray}
where the couplings $g_{\sigma N}~\mathrm{and}~g_{\omega N}$ are determined by fitting the saturation density $\rho_0=0.15~\mathrm{fm^{-3}}$ and the binding energy per baryon $E(\rho=\rho_{0})=-15.7~\mathrm{MeV}$ at the saturation point \cite{pandaHybridStarsQuarkmeson2004}. And the remaining coupling $g_{\rho N}$ is fixed by fitting the asymmetry energy coefficient $a_{sym}=32.5~\mathrm{MeV}$ at the saturation density \cite{pandaHybridStarsQuarkmeson2004}.
The meson masses are taken as $m_\sigma=550\mathrm{~MeV},m_\omega=783\mathrm{~MeV}\mathrm{~and~}m_\rho=770\mathrm{~MeV}$. According to the following relations 
\begin{eqnarray}
	g_{\sigma B}=x_{\sigma B}g_{\sigma N},\quad g_{\omega B}=x_{\omega B}g_{\omega N},\quad g_{\rho B}=x_{\rho B}g_{\rho N}\nonumber
\end{eqnarray}
the hyperon-meson coupling constants are determined, with $x_{\sigma}=0.7\mathrm{~and~}x_{\omega}=x_{\rho}=0.783$ \cite{glendenningReconciliationNeutronstarMasses1991,pandaHybridStarsQuarkmeson2004}. The determination of the coupling constant enables us to parametrize the effective mass as a function of $\sigma$ \cite{juHadronquarkMixedPhase2021,whittenburyQuarkmesonCouplingModel2014,guichonQuarkMesonCouplingQMC2018}.

For the QMC model, the meson field equations of motion in the mean-field approximation at Hartree level are given by \cite{pandaHybridStarsQuarkmeson2004,rikovskastoneColdUniformMatter2007}
\begin{eqnarray}
	m_{\sigma}^{2}\sigma&=&\sum_{B}\frac{2S_{B}+1}{2\pi^{2}}\left[-\frac{\partial M_{B}^{*}(\sigma)}{\partial \sigma}\right]\nonumber\\
	&&\times\int_{0}^{k_{B}}\frac{M_{B}^{*}(\sigma)}{\sqrt{k^{2}+M_{B}^{*2}(\sigma)}}k^{2}dk,
	\\
	m_{\omega}^{2}\omega_{0}&=&\sum_{B}g_{\omega B}(2S_{B}+1)k_{B}^{3}/(6\pi^{2}),
	\\
	m_{\rho}^{2}\rho_{03}&=&\sum_{B}g_{\rho B}I_{3B}(2S_{B}+1)k_{B}^{3}/(6\pi^{2}),
\end{eqnarray}
where $S_{B},I_{3B},k_{B}$ are the 
spin, isospin projection, and the Fermi momentum of different species of baryon B, respectively.

The energy density and pressure including leptons can be derived from the Lagrangian density in the mean-field approximation \cite{juHadronquarkMixedPhase2021,chang-qunPhaseTransitionsNeutron2008,carrollPhaseTransitionQuarkmeson2009,pandaHybridStarsQuarkmeson2004}
\begin{eqnarray}
 	\varepsilon_{\mathrm{HP}}&=&\sum_{B}\frac{2S_{B}+1}{2\pi^{2}}\int_{0}^{k_{B}}k^{2}dk\sqrt{k^{2}+M_{B}^{*2}(\sigma)}\nonumber
 	\\
 	&&+\frac{1}{2}m_{o}^{2}\sigma^{2}+\frac{1}{2}m_{o}^{2}\omega_{0}^{2}+\frac{1}{2}m_{\rho}^{2}\rho_{03}^{2}\nonumber
 	\\
 	&&+\sum_{l=e,\mu}\frac{1}{\pi^{2}}\int_{0}^{k_{l}}k^{2}dk\sqrt{k^{2}+m_{l}^{2}},
 	\\
 	P_{\mathrm{HP}}&=& \frac13\sum_{B}\frac{2S_{B}+1}{2\pi^{2}}\int_{0}^{k_{B}}\frac{k^{4}dk}{\sqrt{k^{2}+M_{B}^{*2}(\sigma)}}\nonumber
 	\\
 	&&-\frac{1}{2}m_{\sigma}^{2}\sigma^{2}+\frac{1}{2}m_{\omega}^{2}\omega_{0}^{2}+\frac{1}{2}m_{\rho}^{2}\rho_{03}^{2}\nonumber
 	\\
 	&&+\frac{1}{3}\sum_{l=e,\mu}\frac{1}{\pi^{2}}\int_{0}^{k_{l}}\frac{k^{4}dk}{\sqrt{k^{2}+m_{l}^{2}}}.
\end{eqnarray}

Neutron star matter with baryons and leptons must satisfy the following conditions:
\begin{eqnarray}
	&&\mu_{p}=\mu_{\Sigma^{+}}=\mu_{n}-\mu_{e},\label{eq13}\\
	&&\mu_{\Lambda}=\mu_{\Sigma^{0}}=\mu_{\Xi^{0}}=\mu_{n},\label{eq14}\\
	&&\mu_{\Sigma^{-}}=\mu_{\Xi^{-}}=\mu_{n}+\mu_{e},\label{eq15}\\
	&&\mu_{\mu}=\mu_{e},\label{eq16}\\
	&&\rho_p+\rho_{\Sigma^+}=\rho_e+\rho_{\mu^-}+\rho_{\Sigma^-}+\rho_{\Xi^-},\label{eq17}
\end{eqnarray}
where from Eq.~(\ref{eq13}) to Eq.~(\ref{eq16}) are the $\beta$-equilibrium conditions and Eq.~(\ref{eq17}) is the charge charge neutrality condition. The chemical potentials of baryons and leptons in the above conditions have the form:
\begin{eqnarray}
	&&
	\mu_B=\sqrt{k_{B}^2+{M_B^{*2}}}+g_{\omega B}\omega_0+g_{\rho B}I_{3B}\rho_{03},
	\\
	&&
	\mu_l=\sqrt{k_l^2+{m_l^2}}.
\end{eqnarray}
\section{\label{sec3}QUARK MATTER DESCRIPTION}
Similar to the hadron matter in neutron stars, the three-flavor quark matter also needs to fulfill the $\beta$-equilibrium conditions Eq.~(\ref{eq20}) and Eq.~(\ref{eq21}) and charge neutrality condition \cite{weberStrangeQuarkMatter2005,glendenningPhaseTransitionsCrystalline2001} Eq.~(\ref{eq22}):
\begin{eqnarray}
	&&
	\mu_s=\mu_d=\mu_u+\mu_e,\label{eq20}\\
	&&
	\mu_\mu=\mu_e,\label{eq21}\\
	&&
	\frac{2}{3}\rho_u-\frac{1}{3}(\rho_d+\rho_s)-\rho_e-\rho_\mu=0,\label{eq22}
\end{eqnarray}
which are necessary for calculating the energy density and pressure of the quark matter in hybrid stars. In this section, two different models are used to describe deconfined quark phase. 
\subsection{The MIT bag model}
In the simplest form of the MIT bag model, the quarks are treated as a noninteracting Fermion gas \cite{juHadronquarkMixedPhase2021,chodosNewExtendedModel1974}, then the energy density and pressure of quark matter are expressed as:
\begin{eqnarray}
	\varepsilon_{\mathrm{QP}} &=&\sum_{q=u,d,s}\frac{3}{\pi^{2}}\int_{0}^{k_{q}}k^{2}dk\sqrt{k^{2}+m_{q}^{2}}+B\nonumber\\
	&&
	+\sum_{l=e,\mu}\frac{1}{\pi^{2}}\int_{0}^{k_{l}}k^{2}dk\sqrt{k^{2}+m_{l}^{2}},\\
	P_{\mathrm{QP}} &=&\sum_{q=u,d,s}\frac{1}{\pi^{2}}\int_{0}^{k_{q}}\frac{k^{4}dk}{\sqrt{k^{2}+m_{q}^{2}}}-B\nonumber\\
	&&
	+\frac{1}{3}\sum_{l=e,\mu}\frac1{\pi^{2}}\int_{0}^{k_{l}}\frac{k^{4}dk}{\sqrt{k^{2}+m_{l}^{2}}},
\end{eqnarray}
where $k_q$, $m_q$, and $B$ are the Fermi momentum, quark current masses and the bag constant, respectively. Here the same values of current quark masses are taken as in the QMC model. Note the bag constant $B$ here is a free parameter, unlike that in the QMC model \cite{juHadronquarkMixedPhase2021,maieronHybridStarsColor2004}.
\subsection{The SU(3) NJL Model}
Next we consider the SU(3) NJL model \cite{bernardFlavorMixingDynamical1987,EBERT1986188} with vector interaction to describe quark phase with  the following Lagrangian density \cite{whittenburyHybridStarsUsing2016}:
\begin{eqnarray}
	\mathcal{L}_{\mathrm{NJL}}&=&
	\bar{\psi}\left(i\not\partial-\hat{m}_0\right)\psi\nonumber  
	\\
	&&
	+G_{\mathrm{S}}\sum_{a=0}^{N_{\mathrm{F}}^2-1}\left[\left(\bar{\psi} \lambda_a \psi\right)^2+\left(\bar{\psi} i \gamma_5 \lambda_a \psi\right)^2\right]\nonumber
	\\
	&&
	+K\left\{\det\left[\bar{\psi}(1+\gamma_5) \psi\right]+\det\left[\bar{\psi}(1-\gamma_5) \psi\right]\right\}\nonumber
	\\&&
	-\begin{cases}
		g_\mathrm{V}(\overline{\psi}\gamma^\mu \psi)^2
		\\
		G_{\mathrm{V}}\displaystyle\sum\limits_{a=0}^{N_{\mathrm{F}}^2-1}\left[\left(\bar{\psi} \gamma_\mu \lambda_a \psi\right)^2+\left(\bar{\psi} \gamma_\mu \gamma_5 \lambda_a \psi\right)^2\right]\nonumber
	\end{cases}\\
\end{eqnarray}
where $\hat{m}_0=\mathrm{diag(m_u,m_d,m_s)}\mathrm{~and~}N_{\mathrm{F}}=3$. The term proportional to $G_{\mathrm{S}}$ is a $U(3)_{L}\times U(3)_{R}$ symmetric four-fermion interaction, where $\lambda_a,a=1,2,...,8$ are Gell-Mann matrices, i.e, the generators of $SU(3)$ group with $\lambda_{0}=\sqrt{2/3}\mathrm{~I}$. The second line corresponds to the Kobayashi-Maskawa-’t Hooft (KMT) six-fermion interaction Lagrangian density which breaks the $U(1)_{A}$ symmetry \cite{thooftComputationQuantumEffects1976}, where $K$ is the coupling constant of such interaction. In this work, we employ two different types of vector interactions: the one proportional to $g_\mathrm{V}$ which is flavor-independent \cite{masudaHadronQuarkCrossover2013,whittenburyHybridStarsUsing2016,bratovicRoleVectorInteraction2013,lourencoVectorInteractionStrength2012}, while another one proportional to $G_\mathrm{V}$ is flavor-dependent \cite{masudaHadronQuarkCrossover2013,whittenburyHybridStarsUsing2016}.

In this NJL model, the constituent quark masses are given by the gap equation within the mean-field approximation \cite{yangInfluenceHadronicEquation2008}:
\begin{eqnarray}
	m_i^*=m_i^0-4G_\mathrm{S}\sigma_i+2K\sigma_j\sigma_k,
\end{eqnarray}
where index $(i,j,k)$ corresponds to the circular permutation of quark flavor $(u,d,s)$; $\sigma_i=\langle\bar{\psi}_i\psi_i\rangle, (i=u,d,s)$ are quark condensates, which has the form:
\begin{eqnarray}
	\sigma_q=\langle\bar{\psi}_q\psi_q\rangle=-\frac{3}{\pi^2}\int_{k_q}^{\Lambda}\frac{m_q^*}{\sqrt{k^2+m_q^{*2}}}k^2dk,
\end{eqnarray}
where $k_q$ and $\Lambda$ denote the Fermi momentum of quark and momentum cutoff, respectively. From the Lagrangian density above one can obtain the total energy density and pressure of the quark phase described by this model:
\begin{eqnarray}
\varepsilon_{\mathrm{QP}}&=&\sum_{q=u,d,s}\left[-\frac{3}{\pi^2}\int_{k_q}^{\Lambda}\sqrt{k^2+m_q^{*2}}k^2dk\right]\nonumber\\
&&
+2G_\mathrm{S}\left(\sigma_u^2+\sigma_d^2+\sigma_s^2\right)\nonumber\\
&&
-4K\sigma_{u}\sigma_{d}\sigma_{s}\nonumber\\
&&
\begin{cases}
	+g_\mathrm{V}({\rho_u}+{\rho_d}+{\rho_s})^2\nonumber\\
	+2G_\mathrm{V}({\rho_u}^2+{\rho_d}^2+{\rho_s}^2)\nonumber
\end{cases}\\
&&
+\sum_{l=e,\mu}\frac{1}{\pi^{2}}\int_{0}^{k_{l}}\sqrt{k^{2}+m_{l}^{2}}k^{2}dk
-\varepsilon_0,	\nonumber\\
P_{\mathrm{QP}}&=&\sum_{i=u,d,s,e,\mu}\rho_i\mu_i-\varepsilon_{\mathrm{QP}}.
\end{eqnarray}
where $\rho_u,\rho_d\mathrm{~and~}\rho_s$ are the density of quarks. Here in the NJL model we need to introduce $\varepsilon_0$ to ensure zero energy density in the vacuum \cite{yangInfluenceHadronicEquation2008}. In this work the RKH parameter set \cite{rehbergHadronizationSUNambu1996} is employed:  $m_{u}^{0}=m_{d}^{0}=5.5\mathrm{~MeV},m_{s}^{0}=140.7\mathrm{~MeV},\Lambda=602.3\mathrm{~MeV},G_\mathrm{S}\Lambda^{2}=1.835,\mathrm{~and~}K\Lambda^{5}=12.36$, and values of vector couplings are varied from $0 G_\mathrm{S}$ to $1 G_\mathrm{S}$ discretely \cite{lourencoVectorInteractionStrength2012,bratovicRoleVectorInteraction2013,whittenburyHybridStarsUsing2016,masudaHADRONQUARKCROSSOVERMASSIVE2013}.
\section{\label{sec4}PHASE TRANSITION IN HYBRID STARS}
In this section, we study the two types of phase transitions mentioned above, namely,  first-order transition and crossover between hadronic phase and quark phase, using different quark models for comparison.
\subsection{First-order transition}
For first-order transition we only consider Gibbs construction here, which has the criteria:
\begin{eqnarray}
	P_{\mathrm{HP}}=P_{\mathrm{QP}}=P_{\mathrm{MP}},\mu_{B}^{H}=\mu_{B}^{Q}\mathrm{~and~}\mu_{e}^{H}=\mu_{e}^{Q}.
\end{eqnarray}
The charge neutrality condition and baryon number conservation of two different phases need to be replaced by the global conditions as below
\begin{eqnarray}
	&&
	(1-\chi)\rho_{\mathrm{HP}}^c+\chi \rho_{\mathrm{QP}}^c+\sum_{l=e,\mu}\rho_{l}^c=0,\\&&
	(1-\chi)\rho_\mathrm{HP}+\chi \rho_\mathrm{QP}=\rho,
\end{eqnarray}
where $\rho^c\mathrm{~and~}\rho$ denotes the charge density of two phases and total baryon density, respectively, with $\chi$ a parameter ranging from 0 to 1, i.e., from hadronic phase to quark phase with the existence of mixed phase. Consequently, the energy density takes the following form:
\begin{eqnarray}
	\varepsilon=(1-\chi)\varepsilon_{\mathrm{HP}}+\chi\varepsilon_{\mathrm{QP}}.
\end{eqnarray}
\subsubsection{Between the QMC model and the MIT Bag model}
\begin{figure}[h]
	\includegraphics{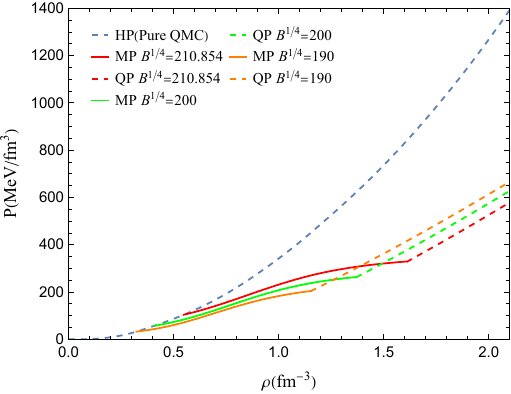}
	\caption{\label{fig1}Pressure as a function of baryon density $\rho$ for hadronic, quark and mixed phases. Blue dashed line represents $P(\rho)$ of pure hadronic phase. The remaining solid and dashed lines of different colors indicate $P(\rho)$ of mixed and quark phase, respectively, with varying bag constant of the MIT bag model,  $B^{1/4}=210.854\mathrm{~MeV}$ (red), $B^{1/4}=200\mathrm{~MeV}$ (green),  $B^{1/4}=190\mathrm{~MeV}$ (orange).}
\end{figure}

\begin{figure*}[t]
	\centering
	\subfloat[\label{fig2a}]{\includegraphics{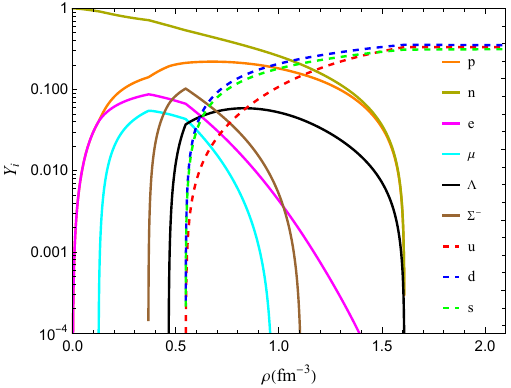}}
	\subfloat[\label{fig2b}]{\includegraphics{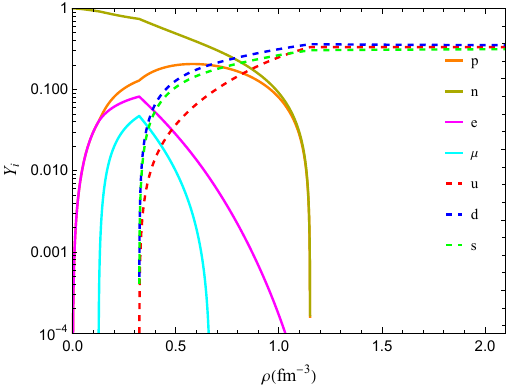}}
	\caption{Relative population of each particle as a function of $\rho$ in generalized beta equilibrium for (a) $B^{1/4}=210.854\mathrm{~MeV}$ and (b) $B^{1/4}=190\mathrm{~MeV}$. The particle species indicated by the different line types and colors are presented in the figure\label{fig2}}
\end{figure*}

The pressure is plotted in Fig.~\ref{fig1} as a function of baryon density $n_B$ with variable bag constants of the MIT Bag model: $B^{1/4}=210.854\mathrm{~MeV},B^{1/4}=200\mathrm{~MeV}\mathrm{~and~}B^{1/4}=190\mathrm{~MeV}$.
As is shown in Fig.~\ref{fig1}, the starting point of mixed phase will move to smaller density as the bag constant decreases, and when the bag constant decreases to a certain value the hyperons may disappear, as is shown by the relative population of each particle in Fig.~\ref{fig2}, where the $Y_i$ denotes the particle fraction: $\rho_i/\rho,(i=N,\Lambda,...,e,\mu,u,d,s)$.
\subsubsection{Between the QMC model and the NJL model}
\begin{figure}[t]
	\centering
	\includegraphics{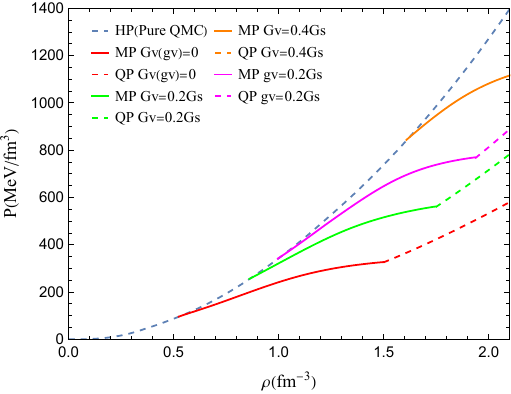}
	\caption{\label{fig3}Pressure as a function of {$\rho$}. Blue dashed line represents $P(\rho)$ of pure hadronic phase. The remaining solid and dashed lines of different colors indicate $P(\rho)$ of mixed and quark phase, respectively, with varying vector couplings of the NJL model, zero vector coupling (red), $G_\mathrm{V}=0.2G_\mathrm{S}$ (green), $G_\mathrm{V}=0.4G_\mathrm{S}$ (orange), $g_\mathrm{V}=0.2G_\mathrm{S}$ (magenta).}
\end{figure}

Similar to the previous part, the pressure with different vector  coupling constants of the NJL model are presented in Fig.~\ref{fig3}. Evidently in the figure, the density at which the mixed phase starts with the same NJL parameter set is strongly correlated to the strength of vector interaction. As the coupling constant increases, the density at which the mixed phase starts increases. Here we wish to note that the flavor-independent type vector interaction can produce a stiffer EOS than the flavor-dependent type EOS when their coupling constants have equal values.

\begin{figure*}[t]
	\centering
	\subfloat[\label{fig4a}]{\includegraphics{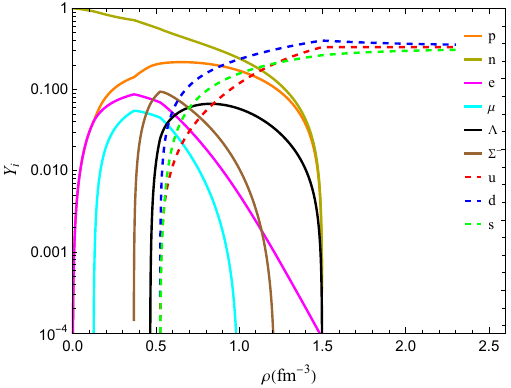}}
	\subfloat[\label{fig4b}]{\includegraphics{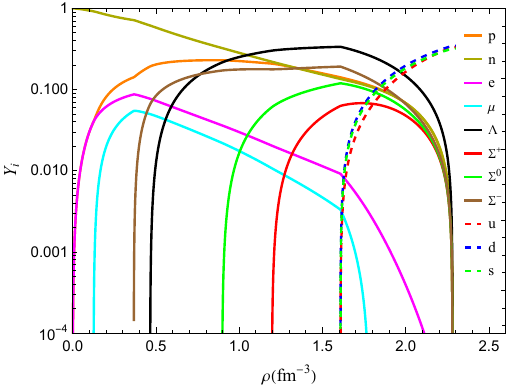}}
	\caption{Relative population of each particle as a function of $\rho$ in generalized beta equilibrium for (a) zero vector coupling and (b) $G_\mathrm{V}=0.4G_\mathrm{S}$\label{fig4}. The particle species indicated by the different line types and colors are presented in the figure.}
\end{figure*}

The relative populations of each particle are displayed in Fig.~\ref{fig4}, which illustrates that the increasing strength of repulsive vector interaction pushes the appearance of quark matter to a higher density.
\subsection{Crossover}
In addition to the first-order phase transition described above, it is also possible to have crossover type of phase transition in hybrid stars \cite{fukushimaPhaseDiagramDense2011,masudaHADRONQUARKCROSSOVERMASSIVE2013}. For crossover, we treat hadrons as finite-sized rather than pointlike objects, so that they start to overlap at some density, and the region where this overlap occurs implies the  beginning of crossover.

In order to obtain crossover type EOSs between hadronic phase and quark phase, we follow the phenomenological $\varepsilon$-interpolation interpolating scheme in Refs.~\cite{masudaHADRONQUARKCROSSOVERMASSIVE2013,masudaHadronQuarkCrossover2013}, which has the form:
\begin{eqnarray}
	&&
	\varepsilon(\rho)=\varepsilon_\mathrm{HP}(\rho)f_-(\rho)+\varepsilon_\mathrm{QP}(\rho)f_+(\rho),\\
	&&
	f_\pm(\rho)=\frac{1}{2}\left[1\pm\tanh\left(\frac{\rho-\bar{\rho}}{\Gamma}\right)\right],
\end{eqnarray}
where $f_\pm(\rho)$ are the interpolating functions and $(\bar{\rho},\Gamma)$ is the parameter set characterizing the center value and the width of the crossover region. With the thermodynamical relation $P(\rho)=\rho^2\frac{\partial(\epsilon/\rho)}{\partial\rho}$ we obtain the pressure:
\begin{eqnarray}
	&&
	P(\rho)=P_\mathrm{HP}(\rho)f_-(\rho)+P_\mathrm{QP}(\rho)f_+(\rho)+\Delta P,\\
	&&
	\Delta P=\rho\left[\varepsilon_\mathrm{HP}(\rho)g_{-}(\rho)+\varepsilon_\mathrm{QP}(\rho)g_{+}(\rho)\right],\\
	&&
	g_{\pm}(\rho)=\pm\frac{2}{\Gamma}\left[e^{\left(\frac{\rho-\bar{\rho}}{\Gamma}\right)}+e^{-\left(\frac{\rho-\bar{\rho}}{\Gamma}\right)}\right]^{-2},
\end{eqnarray}
Note that $\Delta P$ is a correction that guarantees thermodynamic consistency, and its physical meaning is unclear at present stage as it is derived from a phenomenological interpolation, we will discuss both the numerical results with and without the correction term in the following part.

Finally we focus on the choice of parameters $(\bar{\rho},\Gamma)$, which should satisfy two constraints: (1) $dP/d\rho>0$ to ensure the thermodynamic stability of the system, and (2) $\bar{\rho}-2\Gamma>\rho_{0}$ so that the normal nuclear matter will be well described by the hadron EOS \cite{masudaHADRONQUARKCROSSOVERMASSIVE2013,masudaHadronQuarkCrossover2013,masudaHyperonPuzzleHadronquark2016}.
\subsubsection{Between the QMC model and the MIT bag model}
\begin{figure*}[t]
	\centering
	\subfloat[\label{fig5a}]{\includegraphics{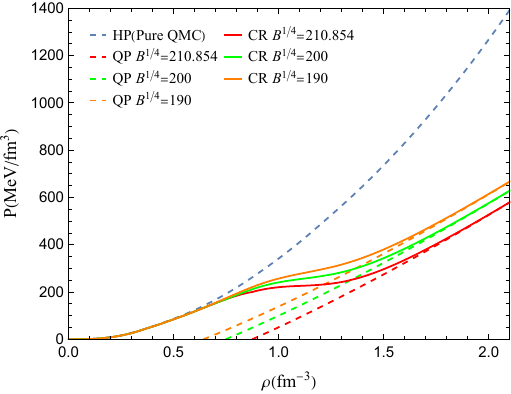}}
	\subfloat[\label{fig5b}]{\includegraphics{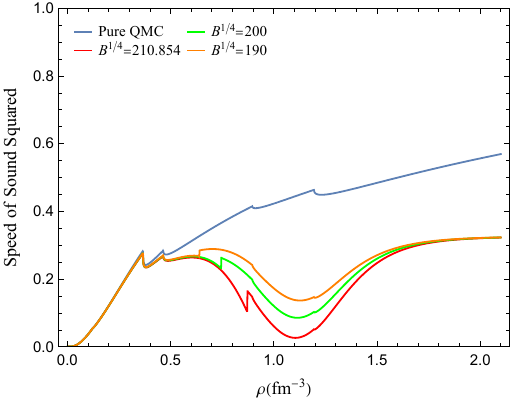}}
	\caption{\label{fig5}(a) Pressure as a function of density {$\rho$}. Dashed curves of different colors indicate $P(\rho)$ of (blue) hadronic or quark phase with varying bag constant, $B^{1/4}=210.854\mathrm{~MeV}$ (red), $B^{1/4}=200\mathrm{~MeV}$ (green), $B^{1/4}=190\mathrm{~MeV}$ (orange). Solid lines are interpolated function and the colors represent the same meaning as the dashed lines. (b) Speed of sound squared as a function of $\rho$. Solid curves in red, green, and orange are obtained from the interpolated EOSs with $B^{1/4}=210.854,210,190\mathrm{~MeV}$, while the blue one corresponds to the QMC EOS. }
\end{figure*}

The interpolation between EOSs of the QMC model and MIT Bag model strongly restricts the selection of parameters in order to satisfy the constraint $dP/d\rho>0$ mentioned above. In Fig.~\ref{fig5}, we only plotted the pressure as a function of baryon density $\rho$ without the thermodynamic correction $\Delta P$ as the correction term strongly reduces the pressure in the transition region and violates the thermodynamic stability of EOSs, where the parameters are taken as $(\bar{\rho},\Gamma)=(7\rho_{0},2\rho_{0})$. And the speed of sound as a function of density is also presented.

As illustrated in the Fig.~\ref{fig5b}, the increasing bag constant softens the generated EOSs, similar to those constructed using Gibbs conditions. It is noteworthy that the speed of sound exhibits abrupt behavior which stiffens the EOSs at the density where the quark equation of state produces a positive pressure, and the definition of $c_s^2$ is as follow:
\begin{eqnarray}
	c_s^2=\frac{dP}{d\varepsilon}.
\end{eqnarray} 
This exotic behavior like a first-order phase transition may suggest that the crossover type phase transitions using the current interpolation scheme are not applicable to models in which quark matter is treated as a noninteracting Fermi gas or weakly interacting quarks.

\subsubsection{Between the QMC model and the NJL model}
\begin{figure*}[t]
	\centering
	\subfloat[\label{fig6a}]{\includegraphics{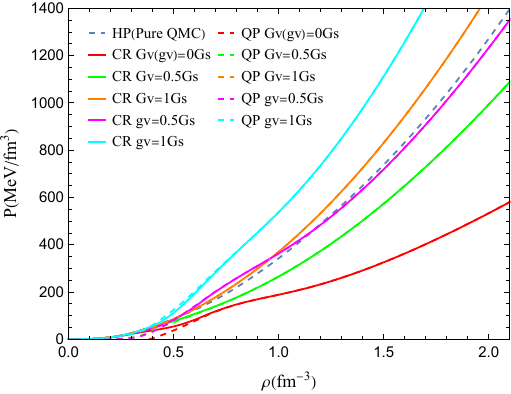}}
	\subfloat[\label{fig6b}]{\includegraphics{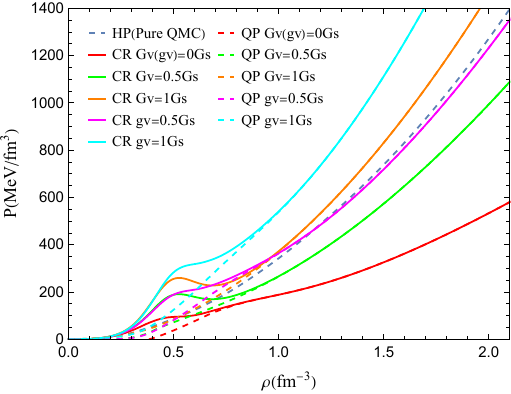}}
	\caption{\label{fig6}Pressure as a function of baryon density {$\rho$}. For plot (a), the interpolation functions do not include the thermodynamic correction $\Delta P$, whereas for those in plot (b) include the correction for thermodynamic consistency. The crossover region is selected as $(3\rho_{0},1\rho_{0})$. The line colors indicate the strength of vector coupling, zero vector coupling (red), $G_\mathrm{V}=0.5G_\mathrm{S}$ (green), $G_\mathrm{V}=1G_\mathrm{S}$ (orange), $g_\mathrm{V}=0.5G_\mathrm{S}$ (magenta), $g_\mathrm{V}=1G_\mathrm{S}$ (cyan). Solid curves are the interpolated functions and colored dashed curves the $P(\rho)$ for quark phase. Blue solid line is again the $P(\rho)$ for hadronic phase.}
\end{figure*}
In the final part of this section, we present the interpolating pressure as a function of baryon density, which the quark phase is described by the NJL model, both with and without $\Delta P$. The interpolating functions are presented in Fig.~\ref{fig6} with the transition region $(\bar{\rho},\Gamma)=(3\rho_{0},\rho_{0})$ \cite{masudaHadronQuarkCrossover2013}. As demonstrated in the figure, the correction term significantly affects EOSs in the transition region, e.g. destabilizes the EOSs with flavor-dependent vector interaction in the right panel . Furthermore, $\Delta P$ may enables EOSs to violate the causality even though they satisfy the thermodynamic stability, which means the speed of sound $c_s^2$ exceeds the speed of light $(c=1)$.

Comparing curves with or without correction term $\Delta P$, one can also see that the type of vector interaction influences the stiffness of EOSs to some extent, which may significant affects the properties of hybrid stars.

\section{\label{sec5}NEUTRON STAR PROPERTIES}
In this section, we use the two models introduced above to investigate the properties of hybrid stars with different phase transitions inside by solving the Tolman-Oppenheimer-Volkov (TOV) equations:
\begin{eqnarray}
	\frac{dP}{dr}&=&-\frac{G}{r^2}\left(M+4\pi Pr^3\right)\left(\varepsilon+P\right)\left(\frac{1-2GM}{r}\right)^{-1},\nonumber\\
	\frac{dM}{dr}&=&4\pi r^2\varepsilon,
\end{eqnarray}
where $G$ denotes the  gravitational constant, and $r$ is the radial distance from the center. Furthermore, after we obtain the mass-radius relations, the unstable hybrid stars need to be excluded in the analysis for their masses decrease as the central energy density goes up \cite{weberStrangeQuarkMatter2005}.

\subsection{Hybrid stars with mixed phase\label{sec.5.a}}
First we investigate the properties of hybrid stars with mixed phases.
\subsubsection{Using the MIT bag model}
\begin{figure*}[!t]
 	\centering
 	\subfloat[\label{fig7a}]{\includegraphics[width=0.33\textwidth,height=0.21\textheight]{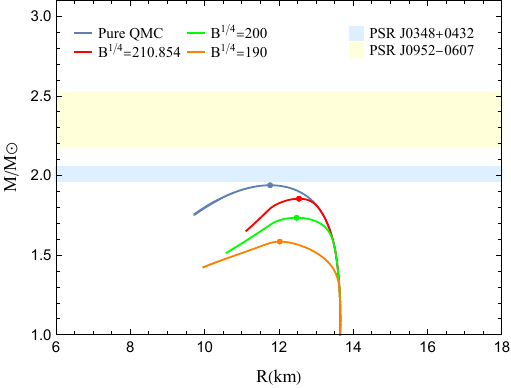}}
 	\subfloat[\label{fig7b}]{\includegraphics[width=0.33\textwidth,height=0.21\textheight]{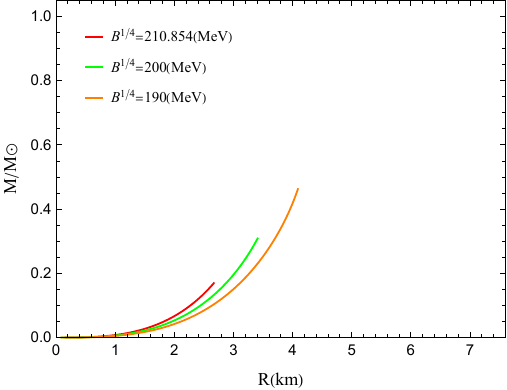}}
 	\subfloat[\label{fig7c}]{\includegraphics[width=0.33\textwidth,height=0.21\textheight]{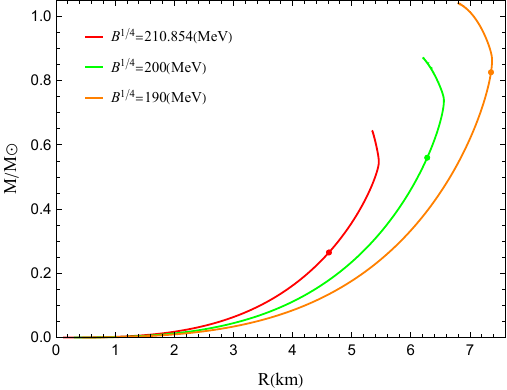}}
 	\caption{\label{fig7}Mass-Radius relations of hybrid stars in panel (a), pure quark cores in panel (b) and mixed core in panel (c) based on the MIT bag model for quark matter. The light yellow band and the light blue band correspond to the mass measurement of PSR J0952-0607 ($2.35\pm0.17M_{\odot}$) \cite{romaniPSRJ095206072022} and PSR J0348+0432 ($2.01\pm0.04M_{\odot}$) \cite{antoniadisMassivePulsarCompact2013}. The different color solid dots in panel (a) and (c) represent the maximally massive stars with different bag constants.}
 \end{figure*}

 The resulting mass-radius relations of hybrid stars and quark cores are presented in Fig.~\ref{fig7}. As shown in Fig.~\ref{fig7a}, the presence of mixed phases and the reduction of the bag constant significantly reduce the maximum mass of the hybrid stars, none of the curves can meet the constraints of the observational data from PSR J0348+0432 \cite{antoniadisMassivePulsarCompact2013} and PSR J0952-0607 \cite{romaniPSRJ095206072022}. The $M-R$ (mass-radius) relations of quark cores in hybrid stars are plotted in Fig.~\ref{fig7b} and Fig.~\ref{fig7c}. For a stable hybrid star, no pure quark matter is found in it even though its mass reaches the maximum value but a inner core with mixed phase. We can also see that the size and mass of such cores increase as the bag constant decrease, then the maximum mass and radius can even reach $0.83M_{\odot}\mathrm{~and~}7.35\mathrm{km}$. Detailed data are presented in Table~\ref{table2}.
\begin{table}[b]
	\caption{The properties of hybrid star containing mixed phase based on the MIT bag model for quark matter with varying bag constant.\label{table2}}
	\renewcommand{\arraystretch}{1.5}
	\begin{ruledtabular}
		\begin{tabular}{ccccc}
			$B^{1/4}(\mathrm{MeV})$&
			$M_{max}/M_{\odot}$&
			$\rho_c(\mathrm{fm^{-3}})$&
			$M_{QC}/M_{\odot}$&
			$R_{QC}(\mathrm{km})$
			\\
			\hline
			210.854&	1.85&	0.82&    0.27&  4.61\\
			200&		1.74&	0.83&    0.56&  6.28\\
			190&		1.59&	0.94&    0.83&  7.35\\
		\end{tabular}
	\end{ruledtabular}
\end{table}

\subsubsection{Using the NJL model}
\begin{figure*}[!ht]
	\centering
	\subfloat[\label{fig8a}]{\includegraphics[width=0.33\textwidth,height=0.21\textheight]{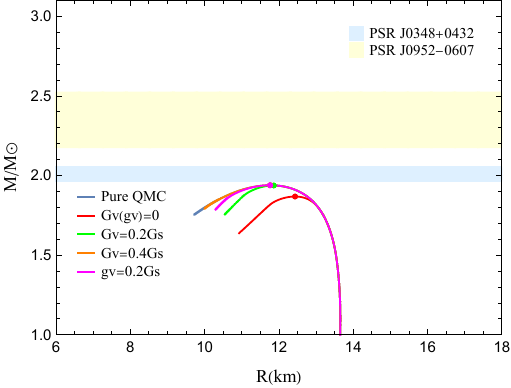}}
	\subfloat[\label{fig8b}]{\includegraphics[width=0.33\textwidth,height=0.21\textheight]{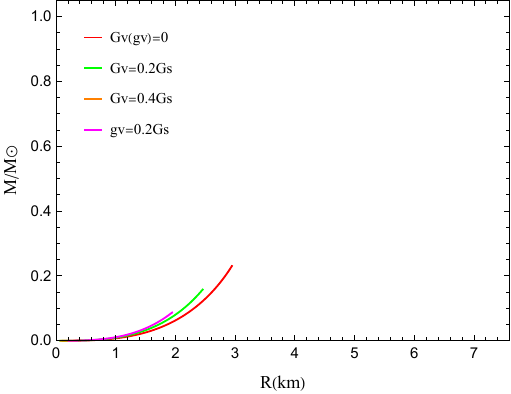}}
	\subfloat[\label{fig8c}]{\includegraphics[width=0.33\textwidth,height=0.21\textheight]{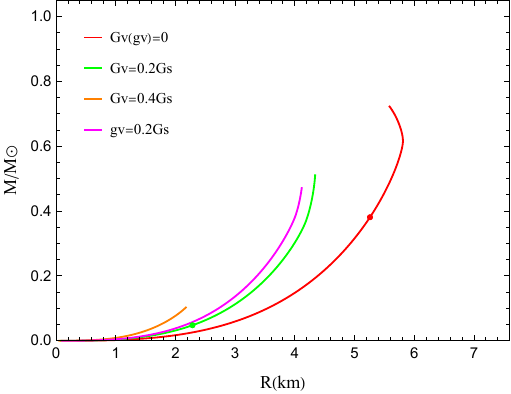}}
	\caption{\label{fig8}Mass-Radius relations of hybrid stars in panel (a), pure quark cores in panel (b) and mixed cores in panel (c) based on the NJL model for quark matter. The different color solid dots in panel (a) and (c) represent the maximally massive stars with varying vector coupling.}
\end{figure*}

The resulting $M-R$ relations are displayed in Fig.~\ref{fig8}. From Fig.~\ref{fig8a}, we are able to analyze the effect that the strength of the vector interaction brings about on the maximum mass of hybrid stars. As the vector coupling constant increases, the maximally massive stars may have greater mass but still under the observational constrains. On the contrary, the cores which contain quark matter shrink with increasing $G_\mathrm{V}\mathrm{~or~}g_\mathrm{V}$, just as we present in Table~\ref{table3}. Moreover, we notice that once the vector coupling constant reaches a relatively large value (such as $G_\mathrm{V}=0.4G_\mathrm{S}\mathrm{~and~}g_\mathrm{V}=0.2G_\mathrm{S}$), quark matter will disappear in a stable hybrid star even reaches the maximum mass, as is shown in Fig.~\ref{fig8c}.
\begin{table}[b]
	\caption{The properties of hybrid star containing mixed phase based on the NJL model for quark matter with varying vector coupling and different vector interaction types. The asterisks (*) in this table indicate that the quark matter does not appear. \label{table3}}
	\renewcommand{\arraystretch}{1.5}
	\begin{ruledtabular}
		\begin{tabular}{cccccc}
			$\mathrm{Type}$&
			$\mathrm{Strength}$&
			$M_{max}/M_{\odot}$&
			$\rho_c(\mathrm{fm^{-3}})$&
			$M_{QC}/M_{\odot}$&
			$R_{QC}(\mathrm{km})$
			\\
			\hline
$G_\mathrm{V}(g_\mathrm{V})$&0$G_\mathrm{S}$&1.869&	0.835& 0.381& 5.26\\
$G_\mathrm{V}$&	0.2$G_\mathrm{S}$&	1.938&	0.944&  0.048&  2.29\\
$G_\mathrm{V}$&	0.4$G_\mathrm{S}$&	1.939&	0.958&  *& *\\
$g_\mathrm{V}$&	0.2$G_\mathrm{S}$&	1.939&	0.958&	*& *\\
		\end{tabular}
	\end{ruledtabular}
\end{table}
\subsection{Hybrid stars with crossover\label{sec.5.b}}
Next we turn to the effects on hybrid stars brought about by crossover type phase transition. Since there is no clear boundary between hadronic matter and quark matter in the crossover situation, we use the polytropic index $\gamma=\mathrm{d}(\ln p)/\mathrm{d}(\ln\epsilon)$ to judge the onset of quark matter \cite{annalaEvidenceQuarkmatterCores2020}. The models describing hadronic matter generically predict $\gamma\approx2.5$ around and above saturation density, while $\gamma=1$ in conformal matter, in this work we employed the approximate rule of  $\gamma\lesssim1.75$ continuously to mark the appearance of quark matter \cite{annalaEvidenceQuarkmatterCores2020}. It should be emphasized that the quark core we discuss in the following presentation refers to the sphere containing quark matter within a hybrid star.
\subsubsection{Using the MIT bag model}
\begin{figure}[!h]
	\centering
	\subfloat{\includegraphics{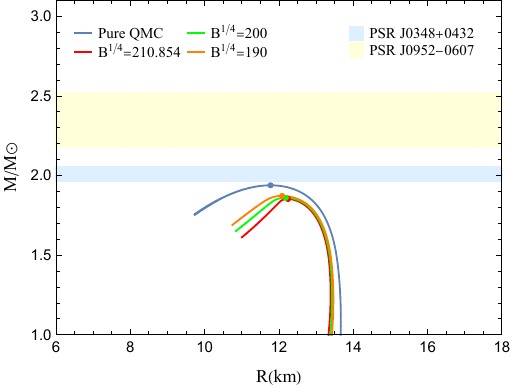}}
	\caption{\label{fig9}Mass-radius relation of hybrid stars based on the interpolation EOSs between the QMC model and the MIT bag model. The crossover region is chosen to be $(7\rho_{0},2\rho_{0})$. The different color solid dots represent the maximally massive stars with different bag constants.}
\end{figure}
Despite the appearance of unreasonable leaps in the EOSs, we still calculate the $M-R$ curves. The properties of hybrid stars, which quark matter is described by the MIT bag model, are shown in Fig.~\ref{fig9} and Table~\ref{table4}. Due to the previously mentioned restrictions on the interpolated EOSs, here we only present the curves obtained with parameters $(\bar{\rho},\Gamma)=(7\rho_{0},2\rho_{0})$. As one can see, the maximum mass of hybrid stars still stay below that of neutron stars containing only hadronic matter described by the QMC model. Nevertheless, the decreasing of bag constant leads to a larger maximum mass, contrary to the case of the first-order phase transition.

\begin{table}[h]
	\caption{The properties of hybrid star in the percolation picture based on the MIT bag model for quark matter with varying vector coupling, and the crossover region is fixed to $(7\rho_{0},2\rho_{0})$. The dashes (—) in this table represent that there is no clear onset for quark matter. \label{table4}}
	\renewcommand{\arraystretch}{1.5}
	\begin{ruledtabular}
		\begin{tabular}{cccccc}
			$B^{1/4}(\mathrm{MeV})$&
			$M_{max}/M_{\odot}$&
			$\rho_c(\mathrm{fm^{-3}})$&
			$M_{QC}/M_{\odot}$&
			$R_{QC}(\mathrm{km})$
			\\
			\hline
		210.854&	1.852&	0.852&   —& —\\
		200&		1.860&	0.866&   —& —\\
		190&		1.872&	0.896&   —& —\\
		\end{tabular}
	\end{ruledtabular}
\end{table}

Another property we need to focus on is the size of the quark core. It must be stressed again that $\gamma\lesssim1.75$ is just a necessary but not a sufficient condition for the appearance of quark matter \cite{somasundaramInvestigatingSignaturesPhase2023}. It turns out that the density at which $\gamma$ crossing 1.75 is far below the crossover window, therefore we cannot have a clear and reliable onset of quark matter with the bag constant range we worked with. This is indicated by the dashes in Table~\ref{table4}.
\subsubsection{Using the NJL model}
\begin{figure}[!t]
	\centering
	\subfloat[\label{fig10a}]{\includegraphics{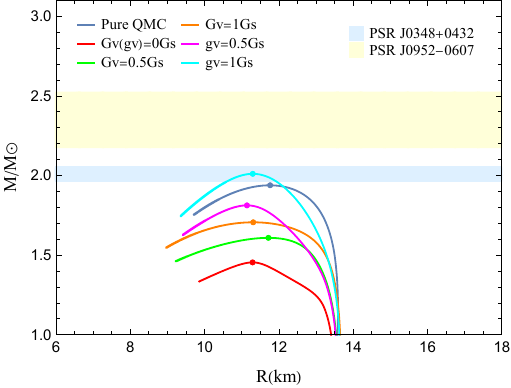}}\\
	\subfloat[\label{fig10b}]{\includegraphics{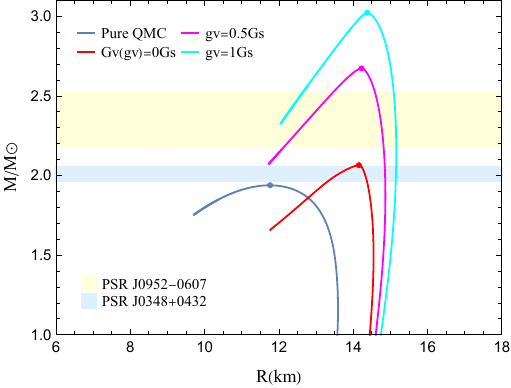}}
	\caption{\label{fig10}Mass-Radius relation of hybrid stars based on the interpolation EOSs between the QMC model and the NJL model. The crossover region is chosen to be $(3\rho_{0},\rho_{0})$. The different color solid dots represent the maximally massive stars with different vector coupling. Plot (a) does not include thermodynamic correction $\Delta P$, whereas plot (b) includes the correction for thermodynamic consistency.}
\end{figure}

Now we display the results related to the NJL model within crossover type phase transition. The resulting $M-R$ curves for several interpolated EOSs, both with and without correction $\Delta P$, are shown in Fig.~\ref{fig10}. It is obvious that when excluding the correction terms, the maximum mass of hybrid meets the constraint from the observational data of pulsar PSR J0384+0342, which is indicated by a light blue band. However, once the correction terms are taken into account, the mass of maximally massive star even exceeds the observational constraint from PSR J0952-0607 and can approach three times solar mass. 

Similar to the situation in the first-order phase transition, the maximum mass of hybrid star is significantly raised by increasing values of vector coupling of the same type. It is also necessary to pay attention to the influence of the central value and width of the phase transition region without $\Delta P$. When $\Gamma$ is fixed and $\bar{\rho}$ is adjusted from $3\rho_{0}$ to $5\rho_{0}$ \cite{masudaHadronQuarkCrossover2013}, the maximum mass increases except for the case $g_\mathrm{V}=1G_\mathrm{S}$. This exception is most likely because the quark phase EOS using vector coupling $g_\mathrm{V}=1G_\mathrm{S}$ is significantly stiffer than the hadronic one at high density. Next fixing $\bar{\rho}$ to $5\rho_{0}$ and changing $\Gamma$ from $1\rho_{0}$ to $2\rho_{0}$, one can observe the decreasing in maximum mass with all values of vector coupling.

Finally, the properties of quark cores are also of interest. Again we employ the $\gamma$ index to determine the onset of quark matter. Note that for interpolation parameters $(5\rho_{0},1\rho_{0})$, a failure in identifying a clear and reliable onset of quark matter also occurs with quite some vector couplings, which is also indicated by the dashes in Table~\ref{table5}. From Table~\ref{table5}, we find that the result presents greater irregularities, since the $\gamma$-dependent criterion used here is only an approximate rule. For maximally massive hybrid stars predicted by EOSs with correction terms and interpolation parameters $(5\rho_{0},2\rho_{0})$, their quark cores generally appears in the interior of the stars at a smaller size and mass comparing to those without corrections. 

\begin{table*}[!t]
	\caption{Hybrid star properties in the percolation picture under variation of crossover region, using the NJL model for quark matter with varying vector coupling and different vector interaction types. The dashes (—) in this table represent that there is no clear onset for quark matter. And the asterisks (*) in this table indicate that a consistent EOS could not be constructed using the chosen interpolation method with variations of vector coupling and interpolation parameters. \label{table5}}
	\renewcommand{\arraystretch}{1.5}
	\begin{ruledtabular}
		\begin{tabular}{ccccccccccc}
		\multicolumn{1}{c}{$(\bar{\rho},\Gamma)$}&
		\multicolumn{1}{c}{$\mathrm{Type}$}&
		\multicolumn{1}{c}{$\mathrm{Strength}$}&
		\multicolumn{2}{c}{$M_{max}/M_{\odot}$}&
		\multicolumn{2}{c}{$\rho_c(\mathrm{fm^{-3}})$}&
		\multicolumn{2}{c}{$M_{QC}/M_{\odot}$}&
		\multicolumn{2}{c}{$R_{QC}(\mathrm{km})$}
		\\
		\cmidrule(r){4-5}
		\cmidrule(r){6-7}
		\cmidrule(r){8-9}
		\cmidrule(r){10-11}
		&&&
		$\Delta P$ &No $\Delta P$&
		$\Delta P$ &No $\Delta P$&
		$\Delta P$ &No $\Delta P$&
		$\Delta P$ &No $\Delta P$
		\\ \hline
	$(3\rho_{0},1\rho_{0})$&$G_\mathrm{V}(g_\mathrm{V})$&0$G_\mathrm{S}$&2.065&1.455&0.610&1.036&0.237&0.208&5.21&3.91\\
	&$G_\mathrm{V}$&	0.5$G_\mathrm{S}$&*&1.609&*&0.867&*&0.762&*&6.83\\
	&$G_\mathrm{V}$&	1$G_\mathrm{S}$&*&1.707&*&0.908&*&1.018&*&7.34\\
	&$g_\mathrm{V}$&	0.5$G_\mathrm{S}$&2.673&1.813&0.631&0.973&0.244&0.326&4.93&4.41\\
	&$g_\mathrm{V}$&	1$G_\mathrm{S}$&3.022&2.011&0.603&0.869&0.242&0.305&4.84&4.27\\
	\hline
	$(5\rho_{0},1\rho_{0})$&$G_\mathrm{V}(g_\mathrm{V})$&0$G_\mathrm{S}$&*&1.859&*&0.830&*&—&*&—\\
	&$G_\mathrm{V}$&	0.5$G_\mathrm{S}$&*&1.872&*&0.790&*&—&*&—\\
	&$G_\mathrm{V}$&	1$G_\mathrm{S}$&*&	1.891&*&0.794&*&—&*&—\\
	&$g_\mathrm{V}$&	0.5$G_\mathrm{S}$&*&1.930&*&0.863&*&—&*&—\\
	&$g_\mathrm{V}$&	1$G_\mathrm{S}$&*&	1.994&*&0.871&*&0.209&*&3.72\\
	\hline
	$(5\rho_{0},2\rho_{0})$&$G_\mathrm{V}(g_\mathrm{V})$&0$G_\mathrm{S}$&*&1.776&*&0.889&*&0.606&*&6.29\\
	&$G_\mathrm{V}$&	0.5$G_\mathrm{S}$&2.267&1.816&0.848&0.829&0.263&0.605&4.35&6.25\\
	&$G_\mathrm{V}$&	1$G_\mathrm{S}$&2.451&1.851&0.818&0.838&0.187&0.735&3.77&6.55\\
	&$g_\mathrm{V}$&	0.5$G_\mathrm{S}$&2.327&1.900&0.876&0.906&0.158&0.384&3.51&4.84
	\\
	&$g_\mathrm{V}$&	1$G_\mathrm{S}$&*&	1.994&*&0.883&*&0.328&*&4.42\\
		\end{tabular}
	\end{ruledtabular}
\end{table*}
\section{\label{sec6}SUMMARY AND OUTLOOK}
In this work, we have investigated the effects of two different phase transitions inside the hybrid star, i.e., the first-order phase transition and crossover, on the maximum mass of hybrid stars and the properties of quark cores inside. Quark matter is described by the MIT bag model and the SU(3) NJL model, respectively, and hadronic matter described by the QMC model.
The Gibbs criterion and interpolation scheme from Ref.~\cite{masudaHADRONQUARKCROSSOVERMASSIVE2013,masudaHyperonPuzzleHadronquark2016} are taken to construct EOSs. There are significant differences in using them to calculate hybrid star properties.

In the first-order phase transition case (Gibbs construction) with the MIT bag model, the maximum mass of the hybrid star increases with the bag constant. The mass and radius of the quark cores decrease as the bag constant increases. With the NJL model, the increasing vector coupling raise the maximum mass of the hybrid star. The quark cores decreases in mass and radius with increasing vector coupling. In neither cases can the maximum mass of hybrid stars exceed that of a pure neutron star and meet observational data. It seems that the appearance of first-order phase transition (Gibbs construction) in a neutron star lowers its maximum mass. And a hybrid star with higher maximum mass may contain a less sizable quark core.

In the crossover transition case with the MIT bag model, the maximum mass of the hybrid star increases as the bag constant decreases, but still fails to meet the observational data. 
The conflict between the crossover region and the index $\gamma$ which fails to give a clear onset for quark matter.
Then in the crossover transition case with the NJL model, the maximum mass of the hybrid star can be significantly boosted to even $3M_{\odot}$ with thermodynamic correction $\Delta P$, much larger than the observational data. In addition, the maximum mass of the hybrid star increases with the vector coupling. The variation of quark cores properties is complicated(Table~\ref{table5}), with or without thermodynamic correction $\Delta P$. It seems that the MIT bag model (without quark interactions) and specific crossover interpolation schemes cannot be used in combination.
The strong vector interactions in the NJL model and thermodynamic correction $\Delta P$ in crossover may be helpful to increase the maximum mass for hybrid stars. 


Our investigation suggest that: In the first-order phase transition case, our result seems to discourage the attempt to increase maximum mass of hybrid stars from quark section; while in the crossover case, the interpolation scheme and quark phase EOSs both have non-negligible influences. Consequently, the choice of models for both hadronic and quark phase still leaves much room for future investigation. For example, the QMC model in the Hartree approximation can be replaced by that in the Hartree-Fock approximation, and the NJL model can be replaced by the PNJL model \cite{FUKUSHIMA2004277,lourencoVectorInteractionStrength2012} or the quasiparticle model \cite{liTidalDeformabilitiesRadii2021,liStrangeQuarkMass2021} for further comparative study. In addition, more work is needed to study the properties of quark cores. Further experimental programs and observational data are expected to provide more inspiration for our future phenomenological study.
\begin{acknowledgments}
	The authors are deeply grateful to the anonymous referee for his/her report that greatly improved the presentation of this manuscript. 
\end{acknowledgments}
\bibliography{yyd}

\end{document}